\documentclass[aps,prb,superscriptaddress,showpacs,twocolumn]{revtex4}%
\usepackage{amsmath}
\usepackage{graphicx}
\usepackage{amsfonts}
\usepackage{amssymb}%
\providecommand{\U}[1]{\protect\rule{.1in}{.1in}}
\begin{document}
\title{Fluctuation-Dissipative Phenomena in a Narrow Superconducting \\Channel Carrying Current Below Critical}
\author{Yu.N.Ovchinnikov}
\affiliation{Max-Plank Institute for Physics of Complex Systems, Dresden, D-01187, Germany }
\affiliation{Landau Institute for Theoretical Physics, RAS, Chernogolovka, Moscow District,
142432 Russia}
\author{A.A.Varlamov}
\affiliation{COHERENTIA-INFM, CNR, Viale del Politecnico 1, I-00133, Rome, Italy}
\date{\today}

\begin{abstract}
The theory of current transport in a narrow superconducting channel accounting
for thermal fluctuations is developed. These fluctuations result in the
appearance of small but finite dissipation in the sample. The value of
\ corresponding voltage is found as the function of temperature (close to
transition temperature $T-T_{\mathrm{c}}$ $\ll T_{\mathrm{c}}$) and bias
current $J<J_{\mathrm{c}}$ ( $J_{\mathrm{c}}$ is a value of critical current
calculated in the framework of the BCS approximation, neglecting thermal
fluctuations). It is demonstrated that the value of the activation energy
$\delta F$ (exponential factor in the Arrenius law) when current approaches to
the critical one is proportional to $\left(  1-J/J_{\mathrm{c}}\right)
^{5/4}.$ This result is in concordance with the one for the affine phenomenon
of the Josephson current decay due to the thermal phase fluctuations, where
the activation energy $\delta F_{J}\sim\left(  1-J/J_{\mathrm{c}}\right)
^{3/2}$(the difference in the exponents is related to the additional current
dependence of the order parameter). Found dependence of the activation energy
on current explains the enormous discrepancy between the theoretically
predicted in Ref. \cite{LA67} and the experimentally observed broadening of
the resistive transition.

\end{abstract}

\maketitle

For the first time the role of fluctuations in the energy dissipation in the
process of current flow through the narrow superconducting channel (NSC) was
considered in the paper of Langer and Ambegaokar \cite{LA67} more than forty
years ago. Publication of this paper has strongly influenced all further
research in this field, it became classical, and corresponding results were
included in multiple monographs and handbooks on superconductivity
\cite{T75,A88,LV04}. Nevertheless, even the authors of Ref. \cite{LA67}
themselves mentioned the striking discrepancy between the predicted and the
experimentally observed \cite{PG67} values for the width of resistive
transition. They attributed such discrepancy to the possible presence of
inhomogeneities in the samples.

Below we will show that point is the Ref. \cite{LA67} contains two incorrect
assumptions which result in the parametrically large overestimation of the
activation energy in the exponent of Arrenius law. The first one is related to
the choice of the form of the free energy functional $F_{s}$, where side by
side with the standard Ginzburg-Landau (GL) part the current-field interaction
term should be taken into account. The second is related to definition of the
saddle point in the Arrenius law. \ The latter should correspond to the second
stationary solution of the GL equation with fixed value of the flowing current
$J$, while the authors of Ref. \cite{LA67} just accepted it in the form
$\Delta\left(  x\right)  =\tanh\left[  x/\left(  \sqrt{2}\xi_{\mathrm{GL}%
}\left(  T\right)  \right)  \right]  ,$ which is correct only in the absence
of current ($\xi_{\mathrm{GL}}\left(  T\right)  $ is the GL coherence length).

In this Letter we have calculated the value of activation energy $\delta F$
for the NSC biased by current $J$. In order to do this we wrote the free
energy functional \ including both GL and the current-field interaction terms,
derived corresponding GL equations, and found the order parameter
$\Delta\left(  x,J\right)  $. Thus we will show that taking into the flowing
current results in the considerable decrease of the value of activation energy
with respect to the result of Ref. \cite{LA67}. For realistic currents this
decrease can reach up to two orders of magnitude.

\emph{Generalities and stability problem}\textit{. }Let us start our
discussion considering the free energy functional written for NSC biased by
current $J$: \textit{ }
\begin{align}
F_{s}  &  =\nu\int d^{3}\mathbf{r}\left\{  \left[  -\tau|\Delta\left(
\mathbf{r}\right)  |^{2}+\frac{\pi\mathcal{D}}{8T}|\partial_{-}\Delta\left(
\mathbf{r}\right)  |^{2}\right.  \right. \label{GLfull}\\
&  \left.  +\frac{7\zeta\left(  3\right)  }{16\pi^{2}T^{2}}|\Delta\left(
\mathbf{r}\right)  |^{4}\right]  +\frac{1}{c}\int d^{3}\mathbf{r}\left(
\mathbf{A}-\left(  c/2e\right)  \mathbf{\nabla}\varphi\right)  \mathbf{\cdot
j}.\nonumber
\end{align}
Here $\nu=mp_{F}/\left(  2\pi^{2}\hbar^{3}\right)  $\ is the density of states
($p_{F}$\ is the electron Fermi momentum), $\tau=1-T/T_{\mathrm{c}}$ is the
reduced temperature, $\partial_{-}=\partial/\partial\mathbf{r-}2ie\mathbf{A/}%
c$, $\zeta\left(  x\right)  $\ is the Riemann zeta-function, $c$ is the speed
of light,
\[
\mathcal{D}=\frac{v_{F}l_{\mathrm{tr}}}{3}\left\{  1+\frac{8T\tau_{tr}}{\pi
}\left[  \psi\left(  \frac{1}{2}\right)  -\psi\left(  \frac{1}{2}+\frac
{1}{4\pi T\tau_{\mathrm{tr}}}\right)  \right]  \right\}
\]
is the diffusion coefficient ($l_{\mathrm{tr}}$ and $\tau_{\mathrm{tr}}$ are
the electron transport mean free path and transport scattering time)\cite{G59}%
, $\psi\left(  x\right)  $ is the Euler psi-function. In order to avoid
cumbersome expressions in intermediate calculations we will use the system of
units where $k_{\mathrm{B}}=1$ and $\hbar=1.$ Nevertheless, in the formulas
important for comparison with experiment we write these constants explicitly.
We assume the current density $j$ to be constant due to the narrowness of the
channel (its cross-section $S\ll\xi_{\mathrm{GL}}^{2}\left(  T\right)  $). It
is the presence of the bias current $J=jS,$ flowing through the
superconducting channel, that results in appearance of the additional
gauge-invariant term in the total free energy functional.

According to general principles, the variation of the free energy functional
(\ref{GLfull}) over modulus $\left\vert \Delta\right\vert $ and
gauge-invariant quantity $\mathbf{\phi}=\mathbf{A}-\left(  c/2e\right)
\mathbf{\nabla}\varphi$ at fixed current leads to GL equations%
\begin{equation}
\left\{
\begin{array}
[c]{c}%
\left[  -\tau-\frac{\pi\mathcal{D}}{8T}\left(  \frac{\partial}{\partial
\mathbf{r}}\mathbf{-}\frac{2ie}{c}\mathbf{A}\right)  ^{2}+\frac{7\zeta\left(
3\right)  }{8\pi^{2}T^{2}}|\Delta|^{2}\right]  \Delta=0\\
-ie\frac{\pi\nu\mathcal{D}}{4T}\left(  \Delta^{\ast}\partial_{-}\Delta
-\Delta\partial_{-}\Delta^{\ast}\right)  =\mathbf{j}%
\end{array}
\right.  . \label{zvezd}%
\end{equation}
Below we will operate in the gauge where $\Delta\left(  x\right)  $ is real
($\varphi$ $=0$).

\begin{figure}[bh]
\includegraphics[width=\columnwidth]{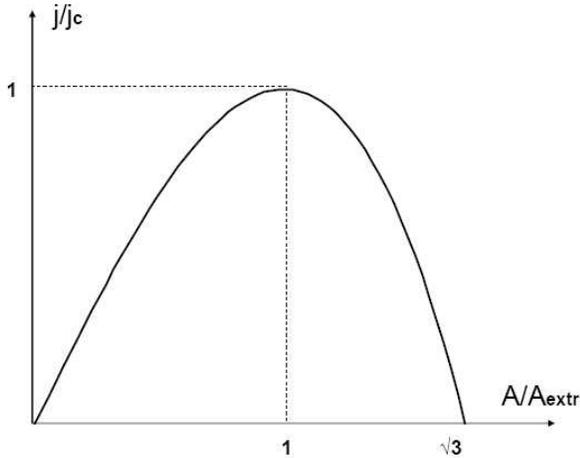}\caption{Schematic
representation of the current density $j$ in the channel as function
of the
vector potential $A$.}%
\label{jAdiagram}%
\end{figure}

The first solution of the system (\ref{zvezd}) with fixed current corresponds
to the the homogeneous state of the NSC with the constant values of the vector
potential $\mathbf{A}$ and the order parameter $\Delta_{0}$ along the
channel:
\begin{equation}
\left\{
\begin{array}
[c]{c}%
\Delta_{0}^{2}{}\left(  A\right)  \mathbf{A=}-\frac{cT}{\pi\nu e^{2}%
\mathcal{D}}\mathbf{j}{}\mathbf{=}const\\
\Delta_{0}^{2}{}\left(  A\right)  =\Delta_{00}^{2}{}(1-\frac{\pi
e^{2}\mathcal{D}}{2Tc^{2}\tau}A^{2})
\end{array}
\right.  , \label{curr}%
\end{equation}
where $\Delta_{00}{}\left(  \tau\right)  =\left[  8\pi^{2}T^{2}\tau/\left(
7\zeta\left(  3\right)  \right)  \right]  ^{1/2}$ is the BCS value of
superconducting order parameter close to critical temperature in the absence
of current. One can see that the current density as the function of vector
potential reaches its maximal value%
\[
j_{\mathrm{c}}=\nu\left(  k_{\mathrm{B}}T\tau\right)  ^{3/2}\frac{16\pi^{5/2}%
}{21\zeta\left(  3\right)  }\sqrt{\frac{2e^{2}\mathcal{D}}{3\hbar}}%
\]
when the vector potential is equal to $|A_{\mathrm{extr}}|=\left[  2Tc^{2}%
\tau/\left(  3\pi e^{2}\mathcal{D}\right)  \right]  ^{1/2}$ (see Fig.
\ref{jAdiagram}).

Let us study the stability of the found homogeneous current state of NSC with
respect to growing current. As is well known, it is determined by the behavior
of the eigenvalues of the operator obtained by linearization of
\textquotedblleft equations of motion\textquotedblright, in our case, Eqs.
(\ref{zvezd}). The state becomes absolutely unstable when the lowest
eigenvalue turns to zero (see, for example, Ref. \cite{O01}). In order to find
corresponding value of the vector potential let us present the functions
$\left\{  \Delta,A\right\}  $ in the form $\left\{  \Delta_{0}+\Delta
_{1},A+A_{1}\right\}  $ assuming $\Delta_{1}\ll\Delta_{0},A_{1}\ll A$ and
\ linearize the system (\ref{zvezd}).

For the fixed current value the corrections $\left\{  \Delta_{1}%
,A_{1}\right\}  $ are connected by the simple relation $A_{1}=-2A\left(
\Delta_{1}/\Delta_{0}\right)  .$\ This relation follows directly from the
first equation of the system (\ref{curr}). Substituting it to the first
equation of \ the system (\ref{zvezd}) one can write the required equation for
the eigenvalue $\lambda:$
\[
\left[  -\tau-\frac{3\pi\mathcal{D}e^{2}}{2Tc^{2}}\mathbf{A}^{2}+\frac
{21\zeta\left(  3\right)  }{8\pi^{2}T^{2}}\Delta_{0}^{2}\left(  A\right)
\right]  \Delta_{1}=\lambda\Delta_{1}.
\]
Remaining in the left-hand side value $\Delta_{0}^{2}$ can be expressed in
terms of \ the vector potential in accordance to Eq. (\ref{curr}), which
results in%

\begin{equation}
\lambda\left(  A\right)  =2\tau-\frac{3\pi\mathcal{D}e^{2}}{Tc^{2}}A^{2}%
=2\tau\left[  1-\left(  \frac{A}{A_{\mathrm{extr}}}\right)  ^{2}\right]  .
\label{lambda}%
\end{equation}
The eigenvalue $\lambda$ becomes zero when the vector potential reaches its
critical value $A=A_{\mathrm{extr}}.$ This is exactly the point of the
absolute instability, where the activation energy in Arrenius law should turn zero.

\emph{Activation energy in decay rate of the NSC.} The system (\ref{zvezd}) at
a given current value has the second, inhomogeneous, solution $\Delta\left(
A,x\right)  $( $x$ is the coordinate along the channel), which determines the
value of activation energy in Arrenius law. In order to find it let us exclude
the vector potential from Eqs. (\ref{zvezd}). One finds%
\begin{equation}
\left(  \frac{\partial\Delta}{\partial x}\right)  ^{2}+\frac{4j^{2}T^{2}}%
{\pi^{2}\nu^{2}e^{2}\mathcal{D}^{2}\Delta^{2}}+\frac{8T\tau}{\pi\mathcal{D}%
}\Delta^{2}-\frac{7\zeta\left(  3\right)  \Delta^{4}}{2\pi^{3}\mathcal{D}T}=C
\label{eq}%
\end{equation}
where $C=const$. Using the dimensionless variables
\begin{equation}
j=j_{\mathrm{c}}\Gamma,\;\Delta^{2}\left(  \Gamma,x\right)  =\Delta_{0}^{2}%
{}\left(  \Gamma\right)  Z\left(  x\right)  ,\;\Delta_{0}^{2}{}\left(
\Gamma\right)  =\Delta_{00}^{2}\mathcal{L}\left(  \Gamma\right)  \emph{,}
\label{dim}%
\end{equation}
one finds from Eqs. (\ref{zvezd}) the cubic equation for $\mathcal{L}$
\begin{equation}
\mathcal{L}^{3}-\mathcal{L}^{2}+\frac{4}{27}\Gamma^{2}=0. \label{L}%
\end{equation}
In the range $A<A_{\mathrm{extr}}$ Eq. (\ref{L}) has the only physically
meaningful solution
\begin{equation}
\mathcal{L}=\frac{1}{3}+\frac{2}{3}\sin\left[  \frac{\pi}{6}+\frac{2}%
{3}\arcsin\sqrt{1-\Gamma^{2}}\right]  . \label{Li}%
\end{equation}

The value
\[
C=\frac{8T\tau\Delta_{0}^{2}{}\left(  \Gamma\right)  }{\pi\mathcal{D}}\left\{
\frac{4\Gamma^{2}}{27\mathcal{L}^{2}}+1-\frac{\mathcal{L}}{2}\right\}
\]
can be found from Eq. (\ref{eq}) by applying the boundary conditions at
infinity: $\Delta\left(  \infty\right)  =0$ and $\partial\Delta/\partial
x|_{x\rightarrow\infty}=0.$

Let us note that the solution of Eq. (\ref{eq}) should be even with respect to
any fixed point $x_{0}:$ $\Delta\left(  x-x_{0}\right)  =\Delta\left(
x_{0}-x\right)  .$ It is why we can assume $x_{0}=0$ and solve Eq. (\ref{eq})
for $x>0$ with a boundary condition$\quad\partial\Delta/\partial x|_{x=0}=0.$
Such solution reads as%
\[
4\sqrt{\frac{T\tau\mathcal{L}}{\pi\mathcal{D}}}x=\int_{2\left(  \mathcal{L}%
^{-1}-1\right)  }^{Z}\frac{dz_{1}}{\left(  1-z_{1}\right)  \sqrt
{z_{1}-2\left(  \mathcal{L}^{-1}-1\right)  }}.
\]
Final integration results in%
\begin{equation}
Z\left(  x\right)  =\frac{1-\mathcal{L}}{\mathcal{L}}+\frac{3\mathcal{L}%
-2}{\mathcal{L}}\tanh^{2}\left(  2x\sqrt{\frac{T\tau\left(  3\mathcal{L}%
-2\right)  }{\pi\mathcal{D}}}\right)  . \label{Z}%
\end{equation}
One can see that corresponding $\Delta\left(  \Gamma,x\right)  $ is reduced to
the one of the Ref. \cite{LA67} only when the flowing current is zero
($\Gamma=0,\mathcal{L}=1$).

Substituting $\Delta^{2}\left(  \Gamma,x\right)  $ found above to Eq.
(\ref{GLfull}) and using Eqs. (\ref{dim})-(\ref{Li}) we obtain the expression
for activation energy $\delta F$%
\begin{align}
\delta F  &  =4\nu S\int_{0}^{\infty}dx\left[  \tau\Delta_{0}^{2}{}\left(
1-Z\right)  -\frac{7\zeta\left(  3\right)  \Delta_{0}^{4}}{16\pi^{2}T^{2}%
}\left(  1-Z^{2}\right)  \right] \nonumber\\
&  -4\nu S\tau\Delta_{0}^{2}{}\left(  1-\mathcal{L}\right)  \int_{0}^{\infty
}dx\left(  \frac{1}{Z}-1\right)  . \label{df}%
\end{align}
Simple integration leads to the final expression for activation energy, valid
for an arbitrary bias current:
\begin{align}
\delta F  &  =2\nu\tau\Delta_{00}^{2}{}\left(  \tau\right)  S\sqrt{\frac
{\pi\mathcal{D}}{T\tau}}\mathcal{L}\left[  \frac{\sqrt{\left(  3\mathcal{L}%
-2\right)  }}{3\mathcal{L}}\right. \nonumber\\
&  \left.  -\frac{\sqrt{1-\mathcal{L}}}{\sqrt{2}}\arctan\left(  \sqrt
{\frac{3\mathcal{L}-2}{2\left(  1-\mathcal{L}\right)  }}\right)  \right]  .
\label{df1}%
\end{align}
When the bias current is close to its critical value one can find from
Eqs.(\ref{Z})-(\ref{df}) that $3\mathcal{L}-2=\frac{2}{\sqrt{3}}\sqrt
{1-\Gamma^{2}}$ and the expression for activation energy is noticeably
simplified:%
\begin{equation}
\delta F=\delta F_{0}\cdot\mathcal{L}\left(  \Gamma\right)  \left[  \frac
{9}{20}\left(  \frac{2}{\sqrt{3}}\sqrt{1-\Gamma^{2}}\right)  ^{5/2}\right]  ,
\label{fff}%
\end{equation}
where
\begin{equation}
\delta F_{0}=\left(  2/3\right)  \nu\tau\Delta_{00}^{2}{}\left(  \tau\right)
S{}\sqrt{\frac{\pi\mathcal{D}}{T\tau}} \label{dfo}%
\end{equation}
is the value of activation energy at zero current ($\Gamma=0$). \

\begin{figure}[b]
\includegraphics[width=\columnwidth]{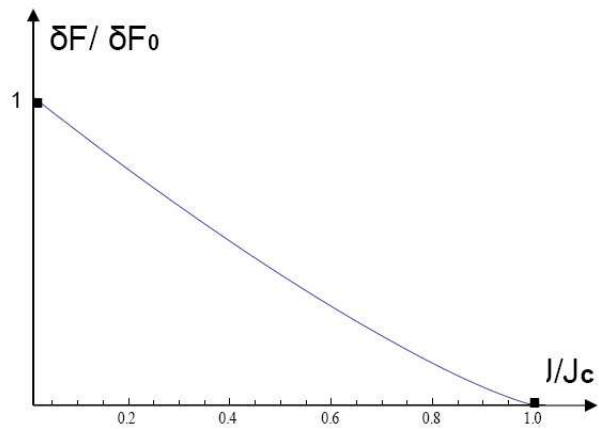}\caption{Current
dependence of the activation energy. $\delta F\left(  J\right)  $}%
\label{comparison}%
\end{figure}

Let us emphasize that in Ref. \cite{LA67} the activation energy in Arrenius
law tends to\textit{ non-zero constant} when current reaches its critical
value. The latter constant differs only by the numerical coefficient of the
order of one from the value\ of activation energy Eq. (\ref{dfo}) calculated
at zero current. At the same time, one can clearly see from Eq. (\ref{fff})
the dramatic effect on $\delta F$ of the correct account for flowing current.
Indeed, the current dependent factor in square brackets strongly depletes the
activation energy Eq. (\ref{fff}) with respect to $\delta F_{0}$. Even not too
close to the critical current when $\sqrt{1-\Gamma^{2}}\approx1/3$ the
activation energy given by Eq. (\ref{fff}) is 30 times smaller then prediction
of the Ref. \cite{LA67}. (see Fig. \ref{comparison}).

Let us indicate the interesting property of the current dependent factor in
the general Eq. (\ref{df1}) exposed by square brackets. In the vicinity of the
critical current two first terms of its Taylor expansion are exactly canceled
out (see Eq. (\ref{fff})). Cancelation of the first term in Eq. (\ref{df1})
can be foreseen and seems trivial, while the second cancelation, which results
in the additional decrease of activation energy $\delta F$ with respect to
$\delta F_{0}$, is surprising.

\emph{Pre-exponential factor.} Let us move to estimation of the
pre-exponential factor in Arrenius law for the number of voltage jumps per
unit time. The GL formalism does not allow its exact definition: in order to
do this it is necessary to know at least the dynamical equations for the order
parameter valid in the wide range of frequencies. Other possibility is to know
the shape of $J-V$ characteristic of the NSC above the critical current, for
$J-J_{\mathrm{c}}\ll J_{\mathrm{c}}.$ Nevertheless, the simplest way to
evaluate the pre-exponential factor is the dimensional analysis which we will
use below.

The Josephson relation Ref. \cite{J62} connects the average voltage $V$ at the
channel to the average time interval $\Delta t$ between the voltage jumps:
$eV=\pi\hbar/\left(  \Delta t\right)  .$ The latter can be estimated as%
\begin{equation}
\Delta t=\left(  \frac{\pi\hbar}{k_{\mathrm{B}}\Delta_{00}\mathcal{L}^{1/2}%
}\right)  \frac{1}{L}\sqrt{\frac{\pi\hbar\mathcal{D}}{k_{\mathrm{B}}%
T\tau\left(  3\mathcal{L}-2\right)  }}\exp\left(  \frac{\delta F}%
{k_{\mathrm{B}}T}\right)  . \label{deltat}%
\end{equation}
Indeed, the first factor should define the characteristic time scale. We
choose it to be in the form $2\pi\hbar/k_{B}\Delta_{0}\left(  \Gamma\right)
.$Then, one should take into account the existence of the \textquotedblleft
zero-mode\textquotedblright, i.e. the arbitrariness of the choice of $x_{0}.$
This means that the instability can arise in an arbitrary point of channel and
it involves the domain of the size of coherence length. This results in
appearance of the second factor in Eq. (\ref{deltat}), which is nothing else
as the ratio of the coherence length in the presence of current to the length
$L$\ of the channel. Finally, accounting\ for the Arrenius exponent we arrive
at the Eq. (\ref{deltat}).

At this point one can write down the J-V characteristics of the NSC close to
transition temperature and for arbitrary current $J<J_{\mathrm{c}}:$
\[
V\left(  J\right)  =\frac{k_{\mathrm{B}}\Delta_{00}\left(  \tau\right)  L}%
{e}\left(  \frac{k_{\mathrm{B}}T\tau\left(  3\mathcal{L}-2\right)
\mathcal{L}}{\pi\hbar\mathcal{D}}\right)  ^{\frac{1}{2}}\exp\left(
-\frac{\delta F}{k_{\mathrm{B}}T}\right)  .
\]

\emph{Discussion.}We demonstrated that the account for the effect of
current flow through the NSC results in a strong suppression of the
energy barrier for the phase slip events with respect to its value
at zero current. In the general Eq. (\ref{df1}) not only the first
term, proportional to $\left( 1-J/J_{\mathrm{c}}\right)  ^{1/4}$,\
but also the next one, proportional to $\left(
1-J/J_{\mathrm{c}}\right)  ^{3/4}$ \ are canceled close to the
critical current $J_{\mathrm{c}}.$ As the result, \ the first
non-vanishing term turns out to be proportional $\left(
1-J/J_{\mathrm{c}}\right)  ^{5/4}$, which is the reason of a strong
reduction of \ the barrier. Moreover, the additional numerical
smallness arises due to the high order of Taylor expansion in Eq.
(\ref{df1}). As a consequence the barrier reduction turns
significant even for currents, being relatively far from the
critical value: for $J_{\mathrm{c}}-J=0.1J_{\mathrm{c}}$ the
reduction factor is 12.4.

One can estimate the width of the temperature smearing $\Delta T$ of
the transition at fixed current $J$ just equating $\delta F\sim
T_{c}.$
In the most interesting case $J\rightarrow J_{c}$ Eq. (\ref{fff}) gives%
\[
\Delta T\left(  J\right)  \sim\frac{\Delta T\left(  J=0\right)  }{\left(
1-J/J_{\mathrm{c}}\right)  ^{5/6}},
\]
where $\Delta T\left(  J=0\right)  \sim Gi_{\left(  1\right)  }$
(Ginzburg-Levanyuk number \cite{LV04} for the NSC) is the width of
transition at zero current. This formula differs from the result of
\ Ref. \cite{LA67} parametrically, by $\left(
1-J/J_{\mathrm{c}}\right)  ^{-5/6}$, which provides the necessary
factor of the order of tens lacking in Ref. \cite{LA67} for the
agreement with the experiment \cite{PG67}.

\begin{figure}[h]
\includegraphics[width=\columnwidth]{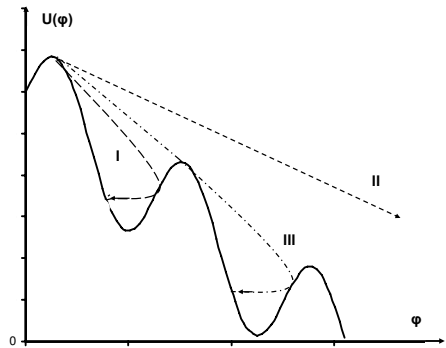}\caption{Different scenarios
for the decay of a metastable state in Josephson junction.}%
\label{wshbrd}%
\end{figure}

Let us mention that the discussed dissipation process is affine to the well
studied phenomenon of the Josephson current decay due to the thermal phase
fluctuations \cite{BP82}. The activation energy for the latter also turns zero
when current, flowing through junction approaches the critical one:%
\[
\delta\emph{F}_{J}\sim\left(  1-J/J_{\mathrm{c}}\right)  ^{3/2}.
\]
The difference between exponents of $\ \delta\emph{F}_{J}$ and
$\delta F$ (Eq. (\ref{fff})  is related to the additional dependence
of the order parameter in NSC on current. This analogy can be useful
when one is trying to understand what kind of the J-V
characteristics one could expect for NSC when currents exceed the
critical value ($J>J_{\mathrm{c}}$)$.$ Three different scenarios are
possible after overcoming the potential barrier in Josephson
junction (see Fig. \ref{wshbrd}) : (I) the system jumps to the
neighbor minimum and for a long time remains around the new minimum
(its phase changes by 2$\pi$); (II) the system switches to the
regime of the \textquotedblleft free over-barrier semiclassical
motion\textquotedblright; (III) the system jumps to any other
minimum with the phase change by 2$\pi N$ ($N$ is an integer number
with some distribution function \cite{M}) and for a long time
remains there.

The realization of this or that scenario depends on the values of the
effective viscosity and $J/J_{\mathrm{c}}$ (depth of the potential well). One
could expect possibility of realization of all this variety of
options\ (depending on $J/J_{\mathrm{c}},T/T_{\mathrm{c}}$ and $l_{\mathrm{tr}%
}$) also in the supercritical regime of the J-V characteristics of the NSC. In
experimental realization the first importance acquires the problem of
overheating related to the very high current densities in superconductor.

The authors acknowledge financial support of the SIMTEC project in the
frameworks of the YII Programme of European Community.

Yu.N.O. is grateful for hospitality of \ \textquotedblleft Tor
Vergata\textquotedblright\ University of Rome (Italy) and grant of
the RFBR.

\end{document}